\documentclass{elsart}
\usepackage{amsmath}
\usepackage{amsfonts}
\usepackage{amssymb}
\usepackage{graphicx}
\newcommand{\wb}{\omega_{\mathrm{b}}}
\newcommand{\nv}{n_{\mathrm{v}}}
\newcommand{\middlefig}{.45\textwidth}
\newcommand{\singlefig}{.75\textwidth}

\begin{document}

\begin{frontmatter}

\title{Influence of moving breathers on vacancies migration}
\date{June 16,2003}

\author[GFNL]{J Cuevas\thanksref{cor}},
\author[GFNL]{C Katerji},
\author[GFNL]{JFR Archilla},
\author[Chris]{JC Eilbeck},
\author[Chris]{FM Russell}
\thanks[cor]{Corresponding author.
             E-mail: jcuevas@us.es}

\address[GFNL]{Grupo de F\'{\i}sica No Lineal.
Departamento de F\'{\i}sica Aplicada I.
ETSI Inform\'{a}tica.
Universidad de Sevilla. Avda. Reina Mercedes, s/n. 41012-Sevilla
(Spain)}

\address[Chris]{Department of Mathematics. Heriot-Watt University.
Edinburgh EH14 4AS (UK)}

\journal{Physics Letters A}

\begin{abstract}

A vacancy defect is described by a Frenkel--Kontorova model
with a discommensuration. This vacancy can migrate when interacts
with a moving breather. We establish that the width of the
interaction potential must be larger than a threshold value in
order that the vacancy can move forward. This value is related to
the existence of a breather centred at the particles adjacent to
the vacancy.

\end{abstract}

\begin{keyword}
Discrete breathers \sep Mobile breathers \sep Intrinsic localized
modes \sep Defects

\PACS  63.20.Pw  
 \sep 63.20.Ry  
 \sep 63.50.+x 
 \sep 66.90.+r 
\end{keyword}

\end{frontmatter}

\section{Introduction}

The interaction of moving localized excitations with defects is
presently a subject of great interest and can be connected with
certain phenomena observed in crystals and biomolecules. Recently,
Sen \emph{et al} \cite{SAR00} have observed that, when a silicon
crystal is irradiated with an ion beam, the defects are pushed
towards the edges of the sample. The authors suggest that mobile
localized excitations called quodons, which are created in atomic
collisions, are responsible for this phenomenon.  The
interpretation is that the quodons are moving discrete breathers
that can appear in 2D and 3D lattices and move following a
quasi-one-dimensional path \cite{MER98}. The interaction of moving
breathers with defects is currently of much interest
\cite{CPAR02b,BSS02,KMN02} (for a review on the concept of
discrete breather see, e.g. \cite{FW98}).

In this paper, we consider a simple one-dimensional model in order
to study how a moving breather can cause a lattice defect to move.
This study is new in the sense that most studies that consider the
interaction of moving discrete breathers with defects, assume that
the position of the latter are fixed and cannot move through the
lattice. The defect that we consider is a lattice
vacancy, which is represented by an empty well or anti-kink in a
Frenkel--Kontorova model \cite{FF96}. The aim of this paper is to
determine in which conditions the vacancy moves towards the ends
of the chain. This is a previous step to reproduce the phenomenon
observed in \cite{SAR00} for higher dimensional lattices.

We have observed, as it will be explained in detail in Section
\ref{sec:numres}, that different phenomena can occur: the vacancy
can move forwards or backwards or remain at rest, and the
breather can be reflected, refracted or trapped. This is quite a
different scenario from the continuous case, in which the vacancy (or
anti-kink) only moves backwards and the breather is always refracted
\cite{KM89}.

\section{The model}

In order to study the migration of vacancies, we consider a
Hamiltonian Frenkel--Kontorova model with anharmonic interaction
potential \cite{BK98}:

\begin{equation}
    H=\sum_n\frac{1}{2}\dot x_n^2+V(x_n)+C\,W(x_{n+1}-x_n).
\end{equation}

\begin{figure}
\begin{center}
    \includegraphics[width=\singlefig]{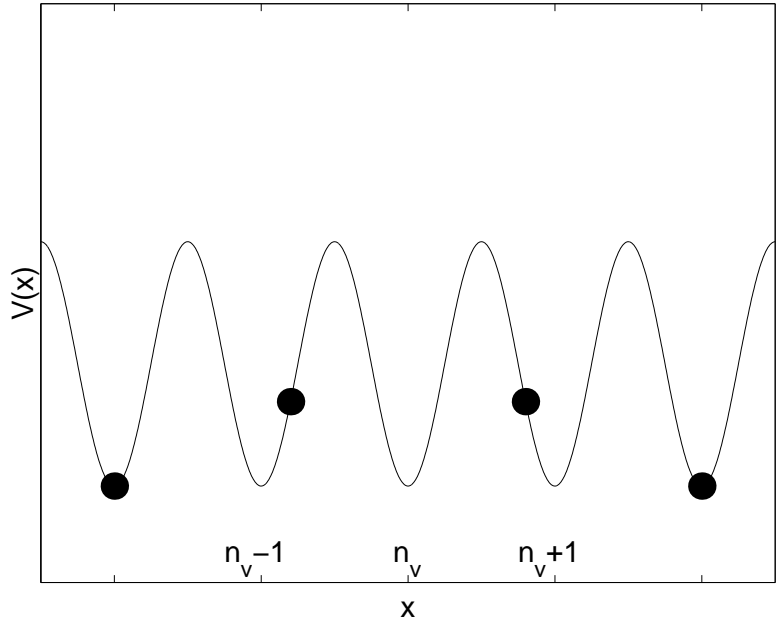}
\caption{Scheme of the Frenkel--Kontorova model with sine-Gordon
on--site potential. The balls represent the particles, which interact
through a Morse potential. The vacancy is located at the site $\nv$.}%
\label{fig:FK}
\end{center}
\end{figure}

The dynamical equations are:

\begin{equation}\label{eq:dyn}
    F(\{x_n\})\equiv \ddot
    x_n+V'(x_n)+C\,[W'(x_n-x_{n-1})-W'(n_{n+1}-x_n)]=0,
\end{equation}%
where $\{x_n\}$ are the absolute coordinates of the particles;
$V(x)$ is the on--site potential, which is chosen of the
sine-Gordon type:

\begin{equation}
    V(x)=\frac{L^2}{4\pi^2}\left(1-\cos\frac{2\pi x}{L}\right),
\end{equation}%
with $L$
being the period of the lattice.
The choice of a periodic potential allows us to
represent  a vacancy easily. Thus, if we denote the vacancy site as
$\nv$ (see figure \ref{fig:FK}), the displacements of the
particles with respect to their equilibrium position are:

\begin{equation}
\left\{ \begin{array}{ll} u_n=x_n-nL & n<\nv \\
\\ u_n=x_n-(n+1)L & n>\nv.\end{array} \right.
\end{equation}

The interaction potential $W(x)$ is of the Morse type:

\begin{equation}
    W(x)=\frac{1}{2b^2}[\exp(-b(x-a))-1]^2,
\end{equation}%
where $a$ is the distance between neighboring minima of the
interaction potential. In order to avoid discommensurations, we
have chosen $L=a=1$. The parameter $b$ is a measure of the inverse
of the width of the interaction potential well. The interaction
between particles is stronger when $b$ decreases. The $1/b^2$
factor allows a Taylor expansion of $W(x)$ at $x=a$ independent on
$b$ up to second order and, in consequence, the curvature at the
bottom of the interaction potential $CW(x)$ depends only on $C$.

The reason for the choice of this potential is twofold. On the one
hand, it represents a way of modelling the interaction
between atoms in a lattice so as the lager the distance between
particles, the weaker the interaction between them becomes. On the
other hand, if a harmonic interaction potential were chosen, apart
 from being unphysical in this model, the
movement of the breather would involve a great amount of phonon
radiation, making it impossible to perform the study developed in
this paper.

Throughout this paper, the results correspond to a breather
frequency $\wb=0.9$. Values of $\wb\in[0.9,1)$ lead to
qualitatively similar results. Values of $\wb\lesssim0.9$ have not
been chosen as moving breathers do not exist \cite{MARIN}.

\section{Numerical Results} \label{sec:numres}

\subsection{Preliminaries}

In order to investigate the migration of vacancies in our model,
we have launched a moving breather towards the vacancy located at the
site $\nv$. This moving breather has been generated using a simplified
form of the marginal mode method \cite{CAT96,AC98}, which consists
basically in adding to the velocity of a stationary breather a
perturbation which breaks its translational symmetry, and letting it
evolve in time. In these simulations, a damping term for the
particles at the edges has been introduced in order that the effects of
the phonon radiation be minimized.

\begin{figure}
\begin{center}
\begin{tabular}{cc}
    \includegraphics[width=\middlefig]{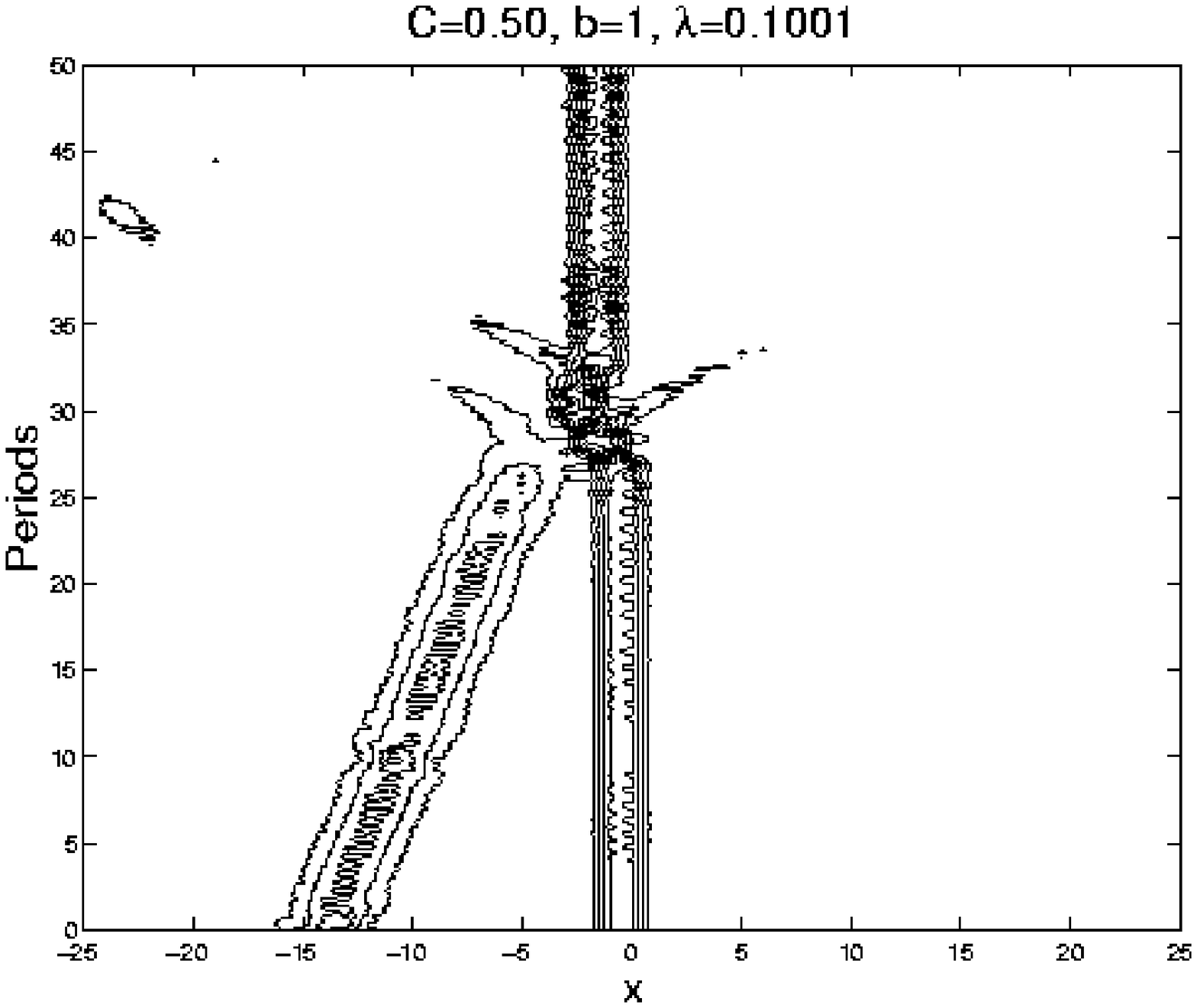} &
   \includegraphics[width=\middlefig]{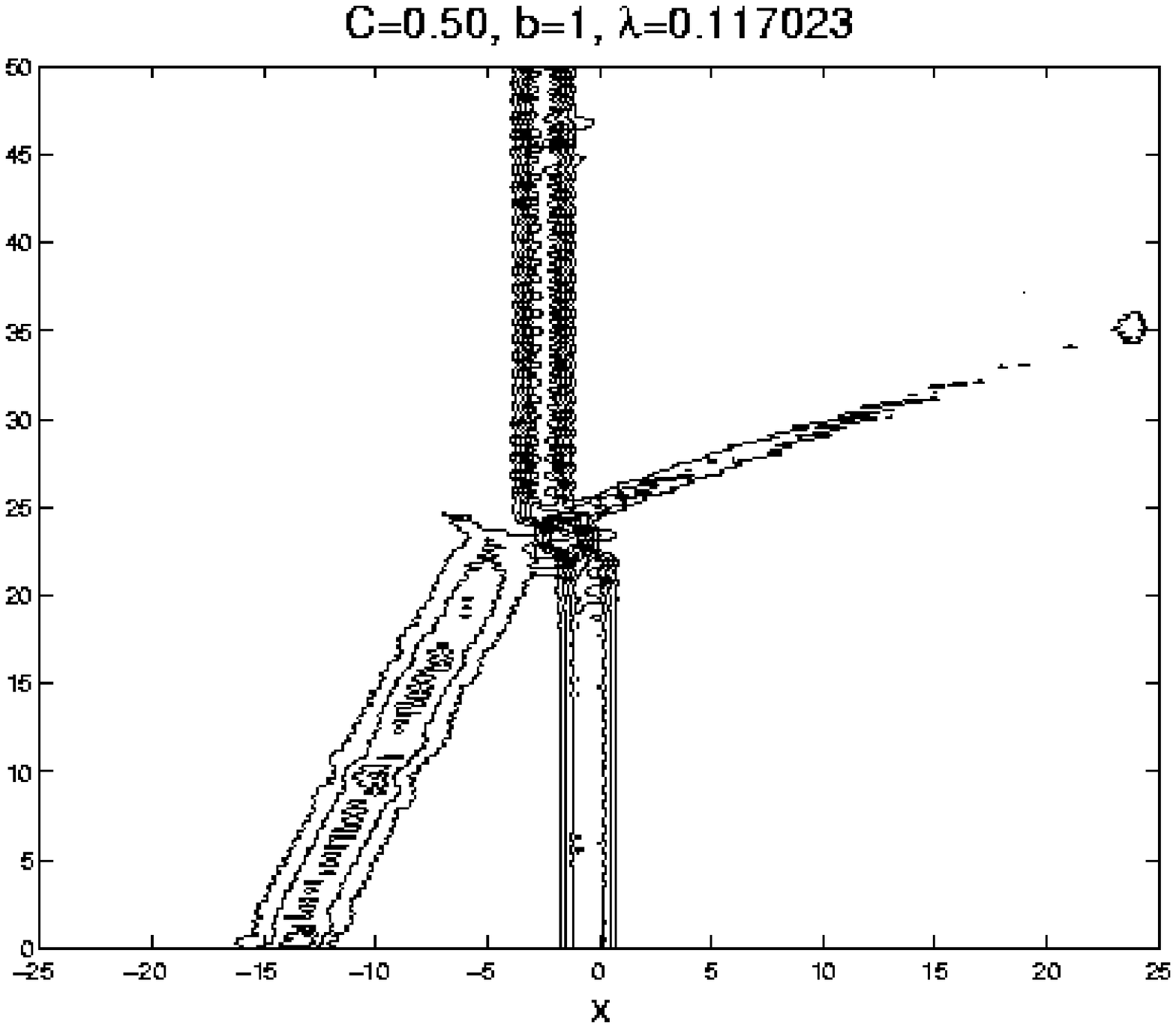}
\end{tabular}
\caption{Energy density plot of the interaction moving
breather--vacancy. The vacancy is located at $\nv=0$. Note that
the vacancy moves backwards (left) and, in the case of the figure
to the right, the breather passes through the vacancy.}%
\label{fig:sim1}
\end{center}
\end{figure}

The initial perturbation, $\{\vec V_n\}$ has been chosen as $\vec
V=\lambda(\ldots,0,-1/\sqrt{2},0,1/\sqrt{2},0,\ldots)$, where the
nonzero values correspond to the neighboring sites of the initial
center of the breather. This choice of the perturbation allows it
to be independent on the parameters of the system  $b$ or $C$.
If the pinning mode were chosen as an initial perturbation, it would
depend on the parameters of the system.

\subsection{Breather--vacancy interaction}

\begin{figure}
\begin{center}
\begin{tabular}{cc}
    \includegraphics[width=\middlefig]{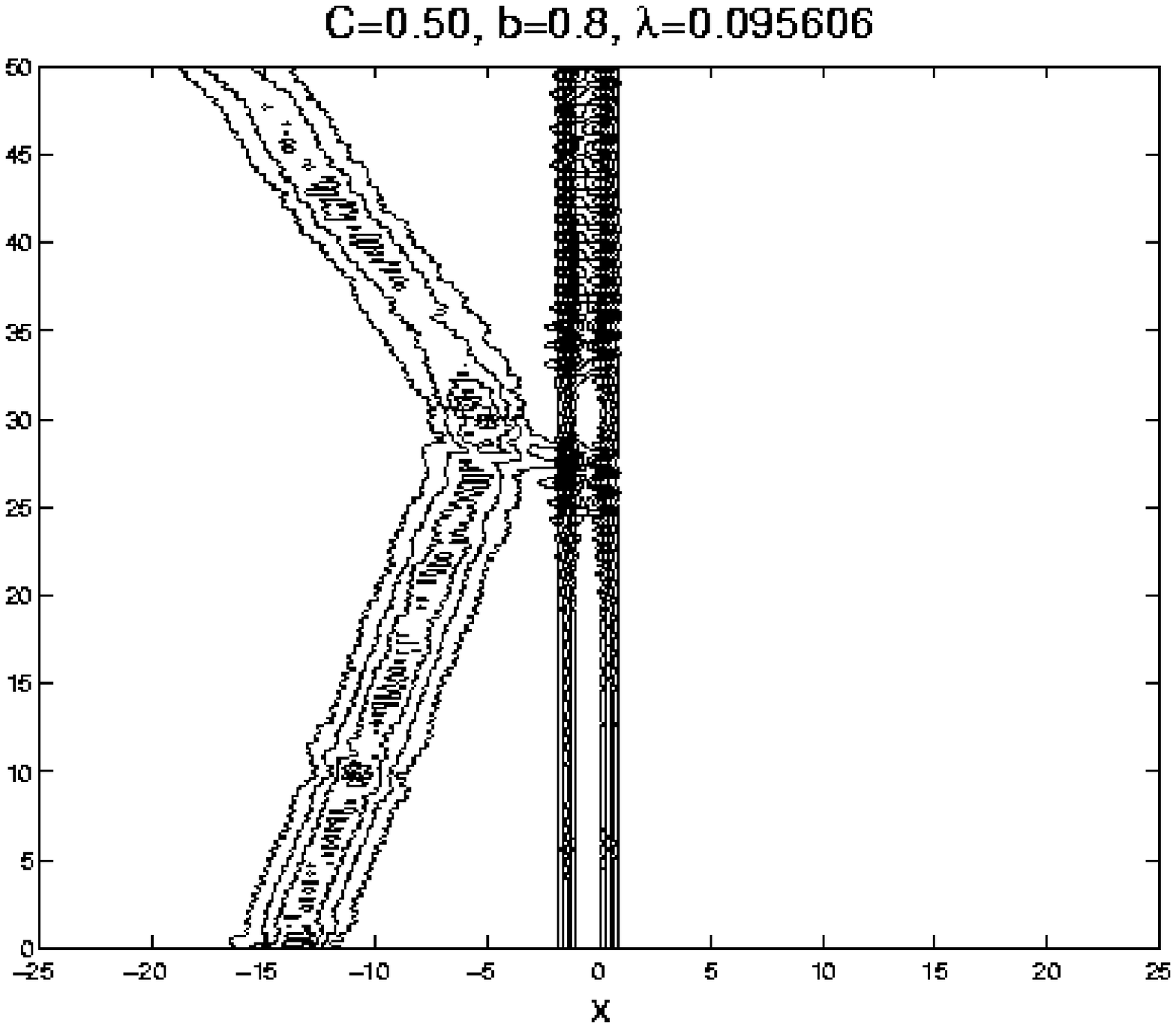} &
    \includegraphics[width=\middlefig]{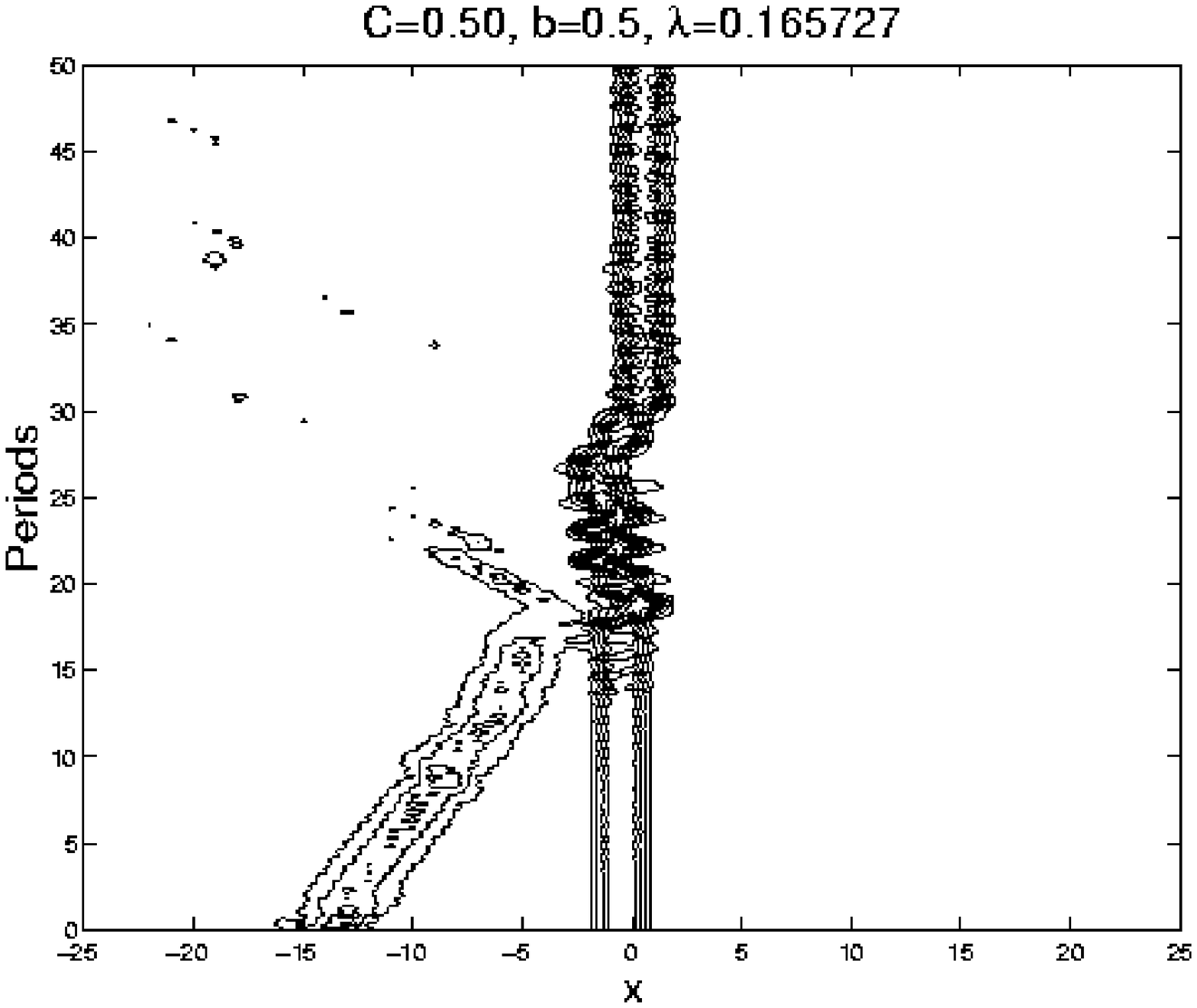}\\
\end{tabular}
\caption{Energy density plot showing the interaction of the moving
breather with the vacancy. The latter is located at $\nv=0$. Note
that, in the figure to the left, the breather is reflected and the
vacancy remains at rest, while in the figure to the right, the
vacancy moves forwards.}%
\label{fig:sim2}
\end{center}
\end{figure}

When a moving breather reaches the site occupied by the
particle adjacent to the vacancy, i.e., the location $\nv-1$, it
can jump to the vacancy site or remain at rest. If the former
takes place, the vacancy moves backwards. However, if the
interaction potential is wide enough, the particle at the $\nv+1$
site, can feel the effect of the moving breather at the $\nv-1$
site and it can also move towards the vacancy site. In the last case,
the vacancy moves forwards. Figures \ref{fig:sim1} and
\ref{fig:sim2} illustrate both phenomena.

\begin{figure}
\begin{center}
\begin{tabular}{cc}
    \includegraphics[width=\middlefig]{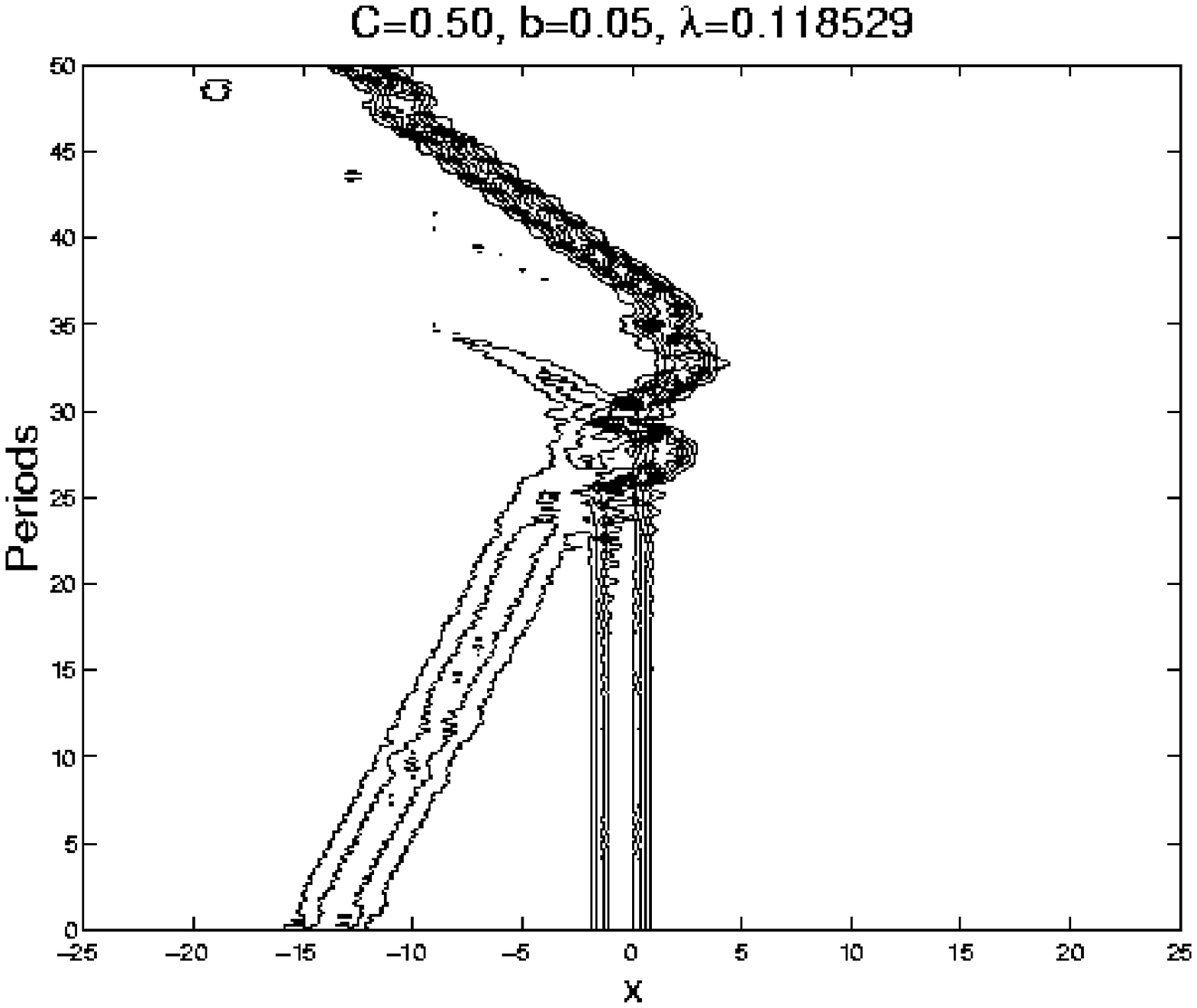}&
    \includegraphics[width=\middlefig]{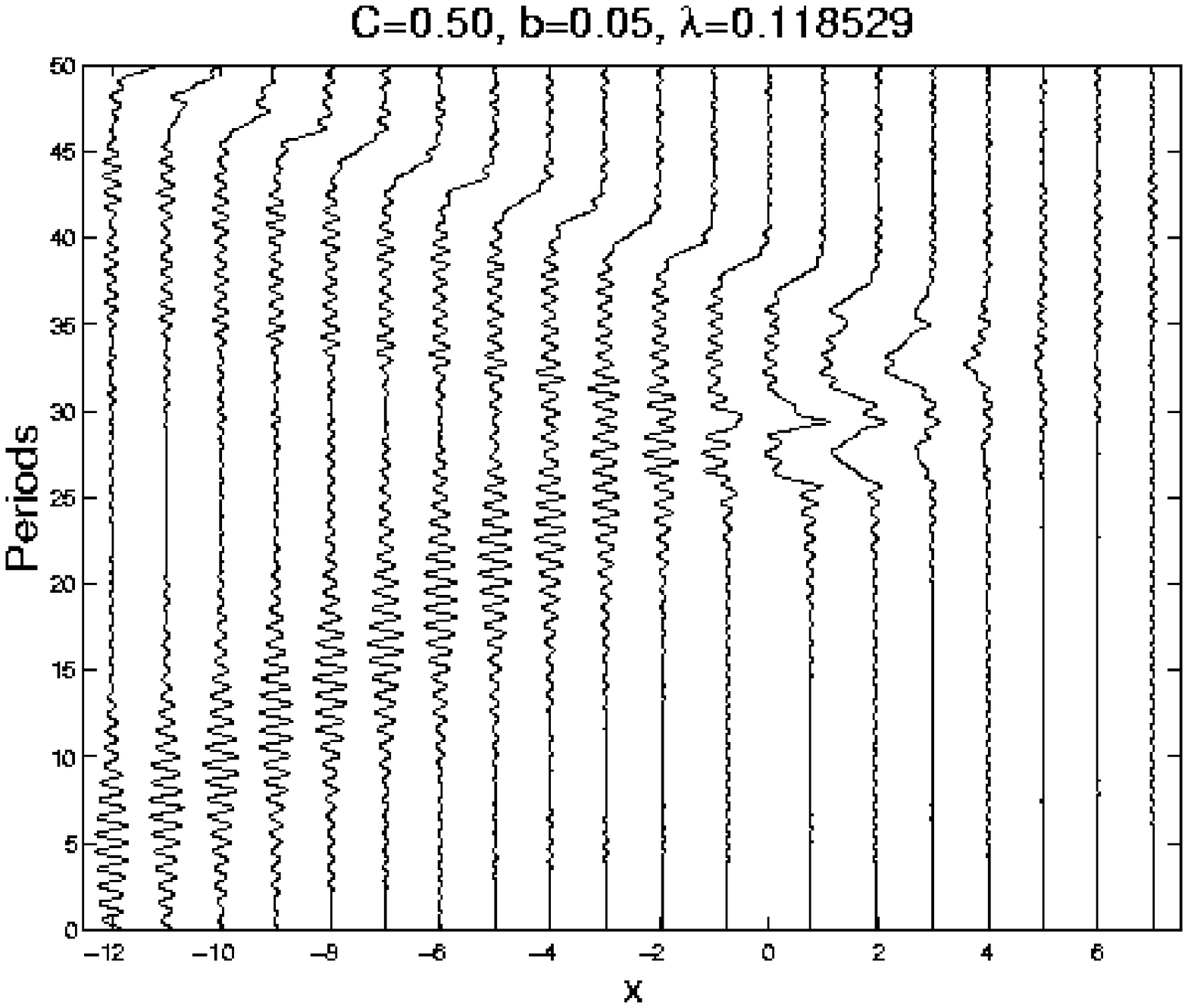}
\end{tabular}
\caption{Left: energy density plot of the interaction moving
breather--vacancy. The vacancy is located at $\nv=0$. It
can travel several sites along the lattice and eventually
stops. Right: detail of the center of the plot showing the variables.}%
\label{fig:sim3}
\end{center}
\end{figure}

It is interesting that the vacancy can migrate along several sites
before stopping if the interaction between particles is strong
enough (see Figure \ref{fig:sim3}). The largest jumps we have
detected are of eleven sites.

\begin{figure}
\begin{center}
\begin{tabular}{cc}
    \includegraphics[width=\middlefig]{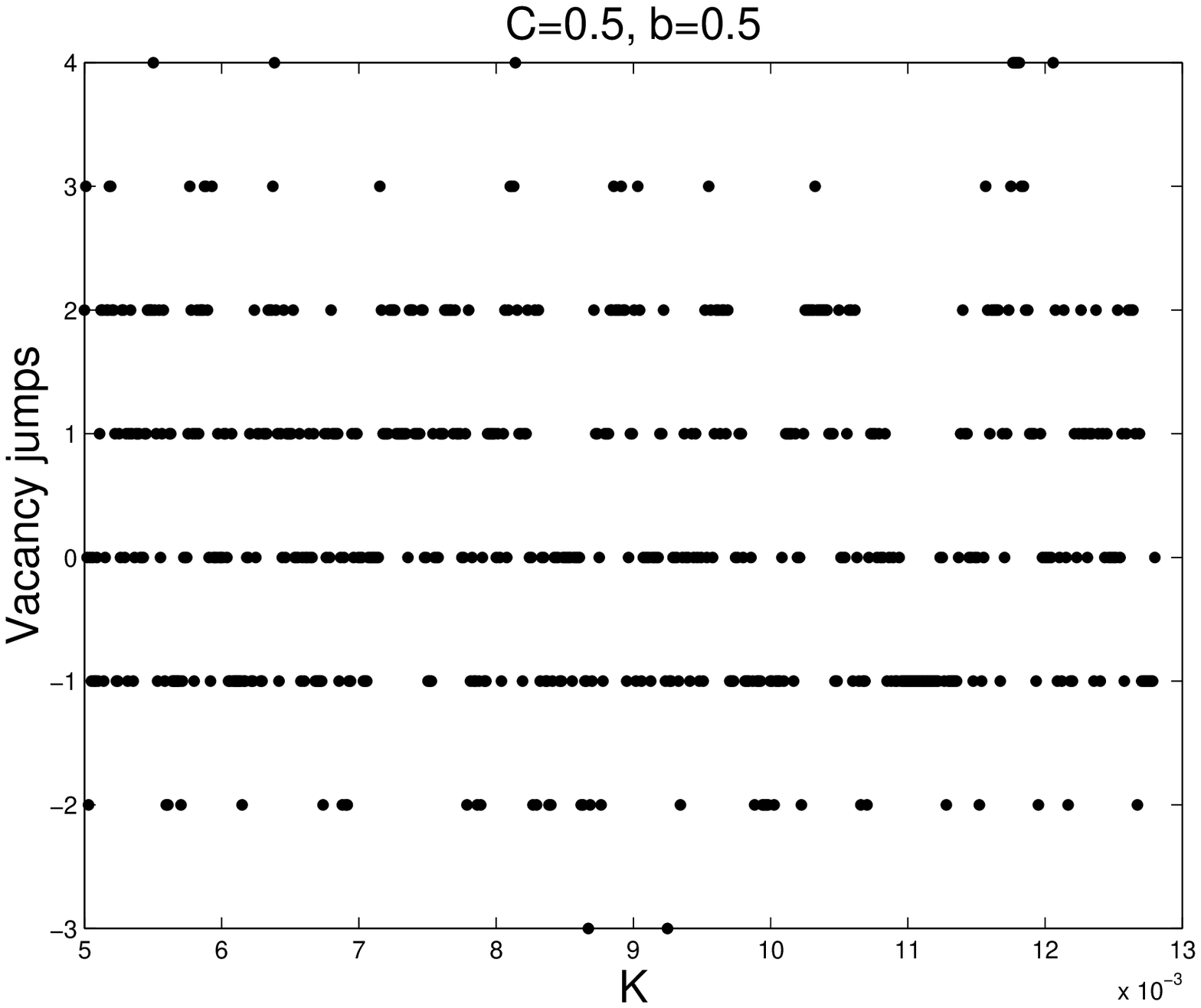} &
    \includegraphics[width=\middlefig]{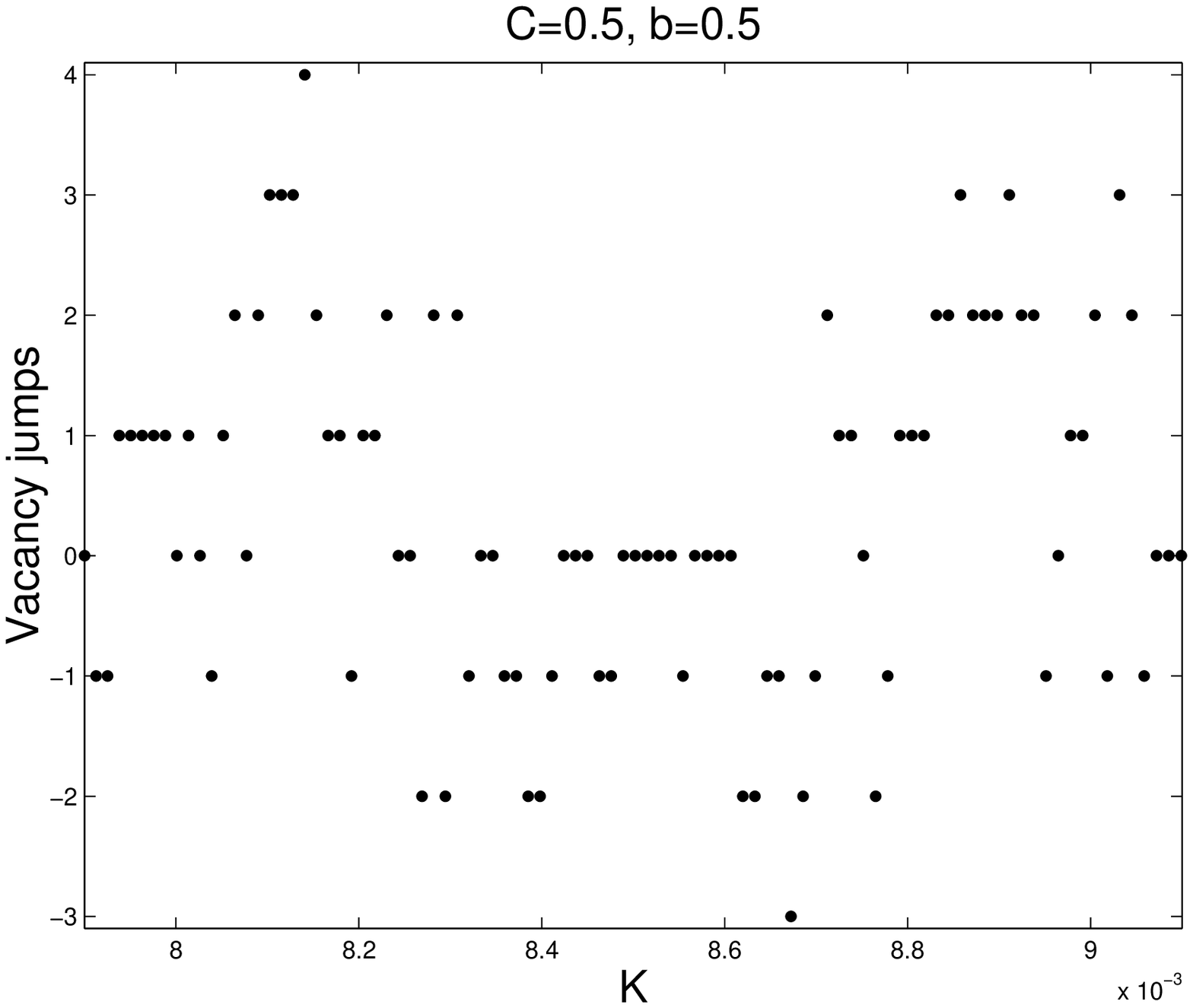}
\end{tabular}
\caption{(Left) Number of sites that the vacancy jumps after its
interaction with a moving breather. (Right) Zoom on a part of the
left figure. Note there is no apparent correlation}%
\label{fig:phase}
\end{center}
\end{figure}

There is no apparent correlation between the characteristics of
the moving breather, e.g. its kinetic energy and its
phase (which has no obvious definition but depends on the initial
distance between the breather and the vacancy and the initial
velocity of the breather). As an example, Figure \ref{fig:phase}
shows the vacancy jumps corresponding to different values of the
translational kinetic energy of the breather. We have not been able to
detect any pattern. The same plot with respect to the breather
distance to the vacancy has a similar appeareance.

\mbox{}\\ \mbox{}

Numerical simulations show that the occurrence of the three
different cases depends highly on the relative phase of the
incoming breather and the particles adjacent to the vacancy.
However, some conclusions can be extracted: 1) The incident
breather always losses energy; 2) The breather can be reflected,
trapped (with emission of energy) or refracted by the vacancy, in
analogy to the interaction moving breather-mass defect
\cite{CPAR02b}; 3) the refraction of the breather (i.e. the
breather passes through the vacancy) can only take place if the
vacancy moves backwards, i.e. the particle to the left jumps one
site in the direction of the breather. A explanation of
this fact is that the particles to the right of the vacancy, in
order to support a moving breather, need a strong interaction which
cannot be provided by the interaction across a vacancy site, because the
distance correspond to the soft part of the Morse potential.

\subsection{Numerical simulations}

As mentioned earlier, the moving breather--vacancy interaction is
highly phase-dependent in a non obvious way. That is, the
interaction depends on the velocity of the breather and the
distance between the breather and the vacancy. Consequently, a
systematic study of the state of the moving breather and the
vacancy after the interaction cannot be performed.

\begin{figure}
\begin{center}
\begin{tabular}{cc}
    \includegraphics[width=\middlefig]{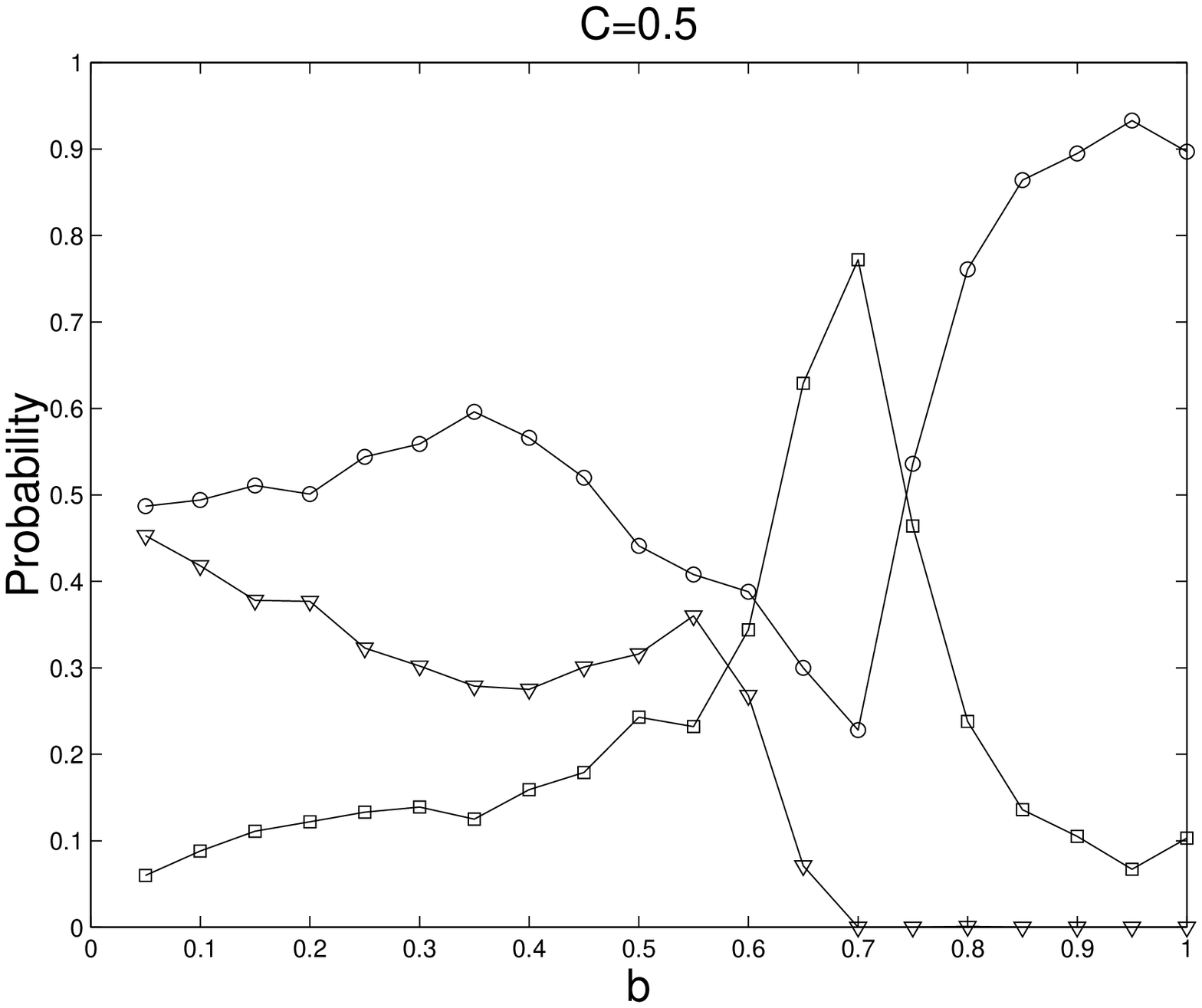} &
    \includegraphics[width=\middlefig]{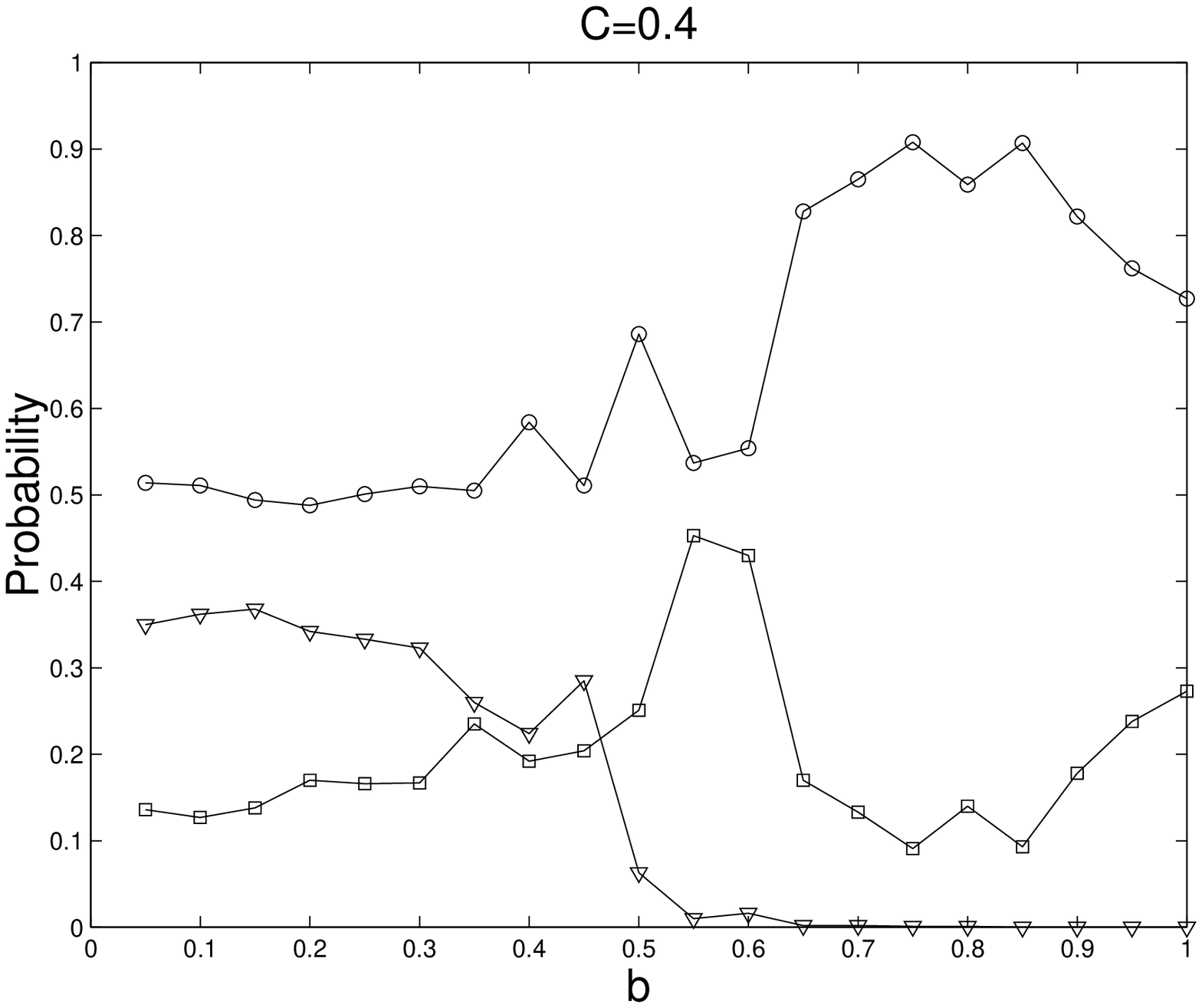}
\end{tabular}
\caption{Probability that the vacancy remains at its site
(squares), moves backwards (circles) or moves forwards
(triangles), for a Gaussian distribution of $\lambda$ as
a function of the inverse potential width $b$.}%
\label{fig:rands}
\end{center}
\end{figure}

\begin{figure}
\begin{center}
\begin{tabular}{cc}
    \includegraphics[width=\middlefig]{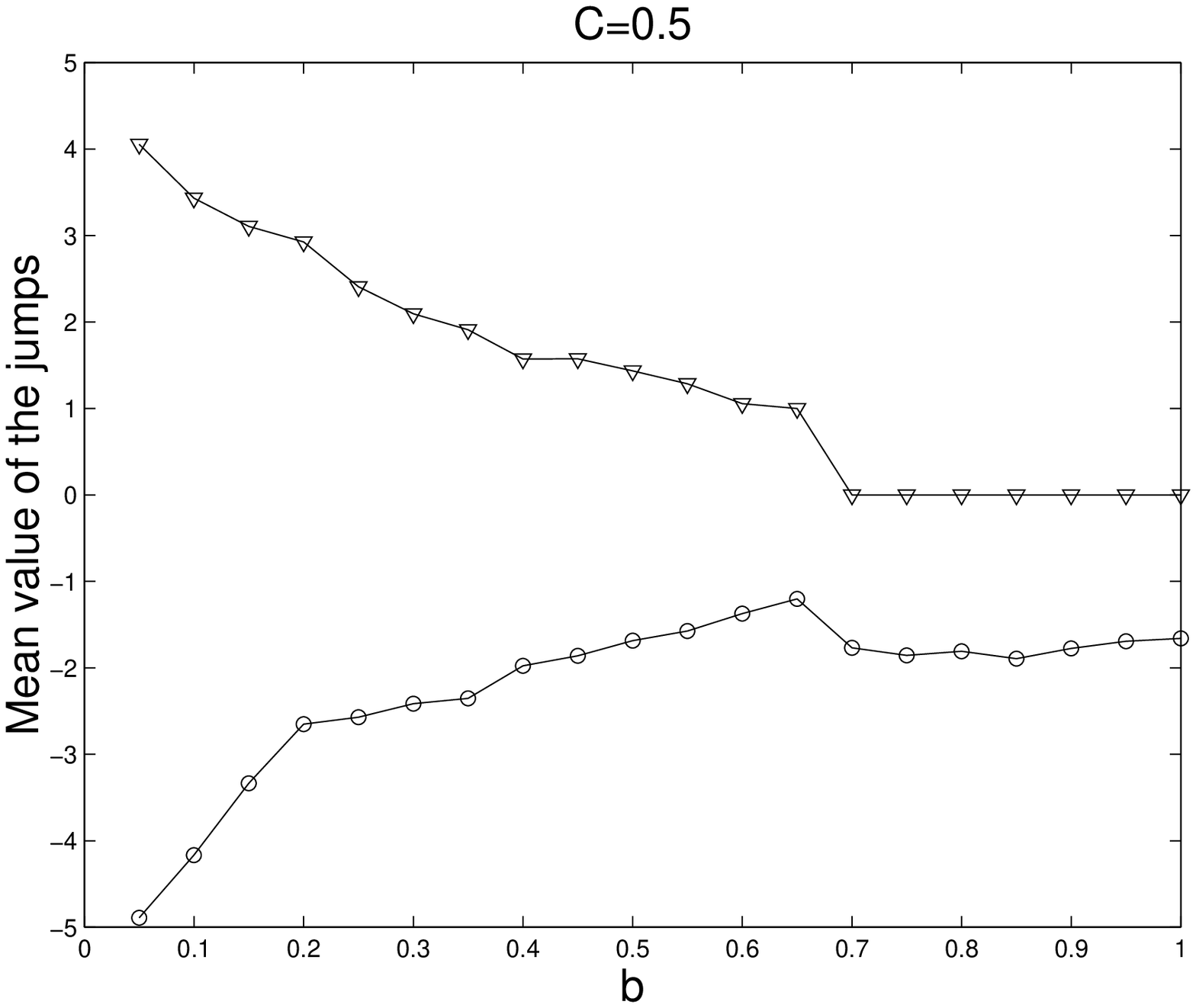} &
    \includegraphics[width=\middlefig]{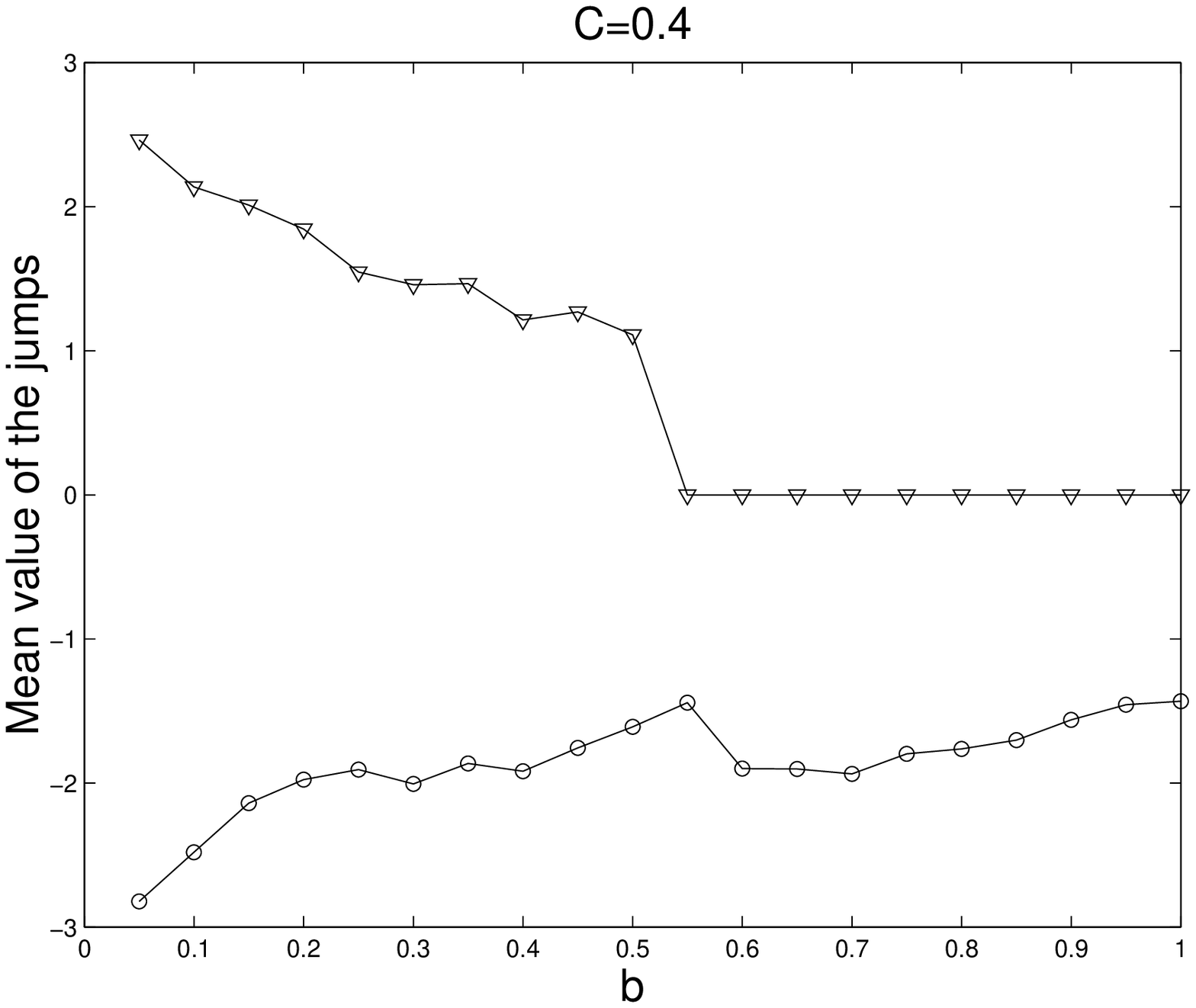}
\end{tabular}
\caption{Mean value of the number of vacancy jumps for the
backward (circles) and forward (triangles) movements as a function
of the inverse potential width $b$. The results correspond to the
simulation performed to obtain Figure \ref{fig:rands}.}%
\label{fig:randav}
\end{center}
\end{figure}

Therefore, we have performed a great number of simulations each
one consisting in launching a single breather towards the vacancy
site. In particular, we have chosen 1000 breathers following a
Gaussian distribution of the perturbation parameter $\lambda$ with
mean value $0.13$ and variance $0.03$ for different values of the
parameters $b$ and $C$. Figure \ref{fig:rands} shows the
probabilities that the vacancy remains at its original site, or
that it jumps backwards or forwards, for $C=0.5$ and $C=0.4$.
Figure \ref{fig:randav} shows the mean values of the number of
vacancy jumps for the forward and backward movement as a function
of the inverse potential width $b$.

An important consequence can be extracted from this figure. There
are two different regions of values for the parameter $b$, separated by a
critical value $b_0(C)$. For $b>b_0(C)$, the probability that the
vacancy moves forwards is almost zero, whereas for $b<b_0(C)$,
this probability is significant. For example, $b_0(C=0.5)\approx0.70$
and $b_0(C=0.4)\approx0.55$.

\begin{figure}
\begin{center}
    \includegraphics[width=\singlefig]{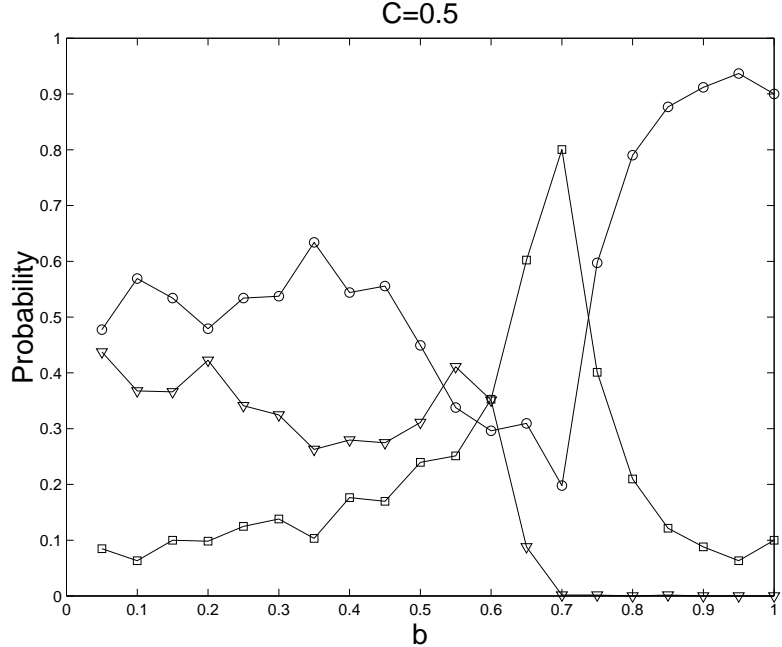}
\caption{Probability that the vacancy remains at its site
(squares), moves backwards (circles) or moves forwards
(triangles), for an uniform distribution of $\lambda$.}%
\label{fig:unifs}
\end{center}
\end{figure}

Figure \ref{fig:unifs} represents this dependence for an uniform
distribution of $\lambda\in(0.10,0.16)$, and shows the occurrence
of the same phenomenon. Thus, this result seems to be independent
on how $\lambda$ is distributed.

\subsection{Analysis of some results. Vacancy breather bifurcation.}

\begin{figure}
\begin{center}
\begin{tabular}{cc}
    \includegraphics[width=\middlefig]{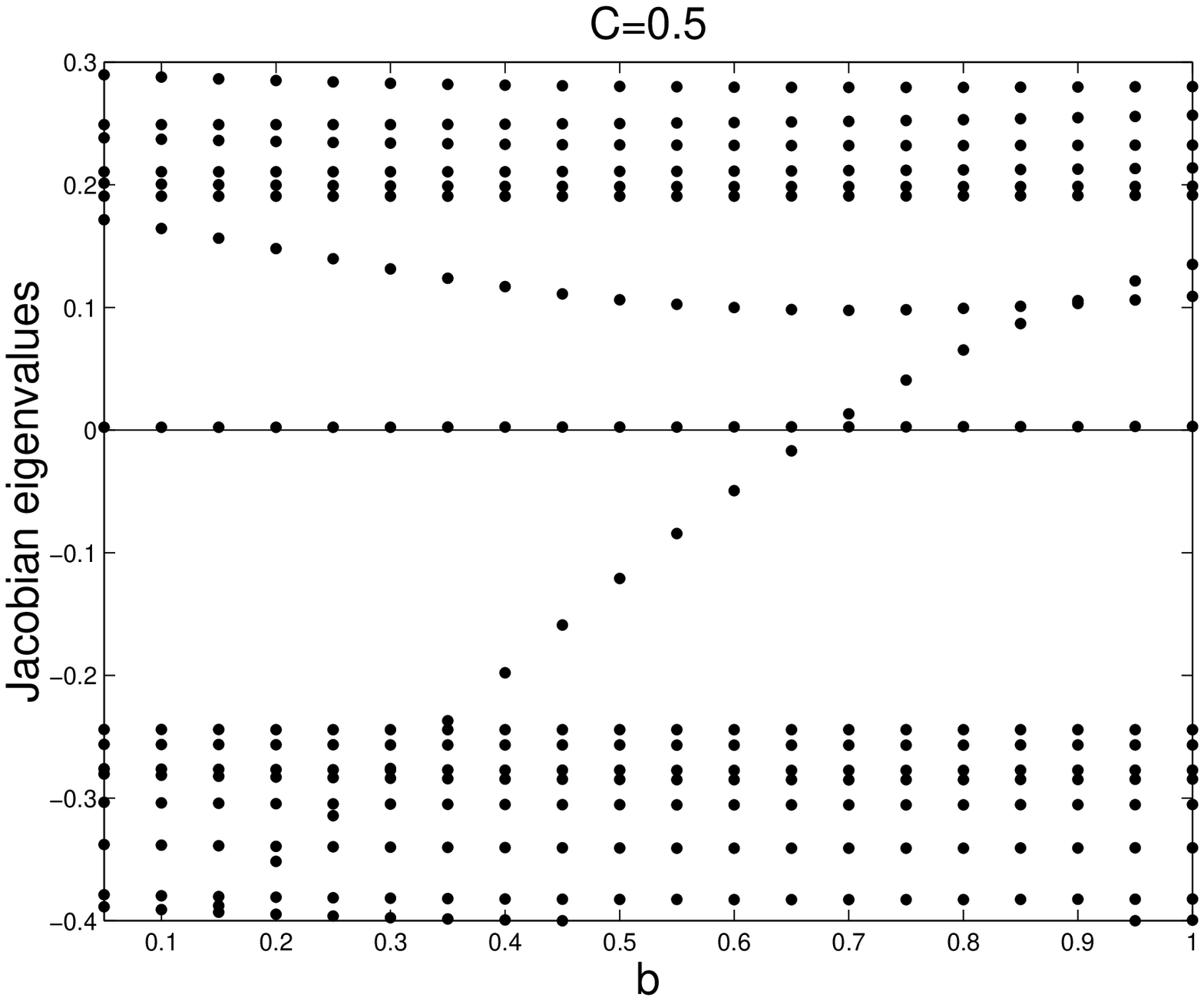} &
    \includegraphics[width=\middlefig]{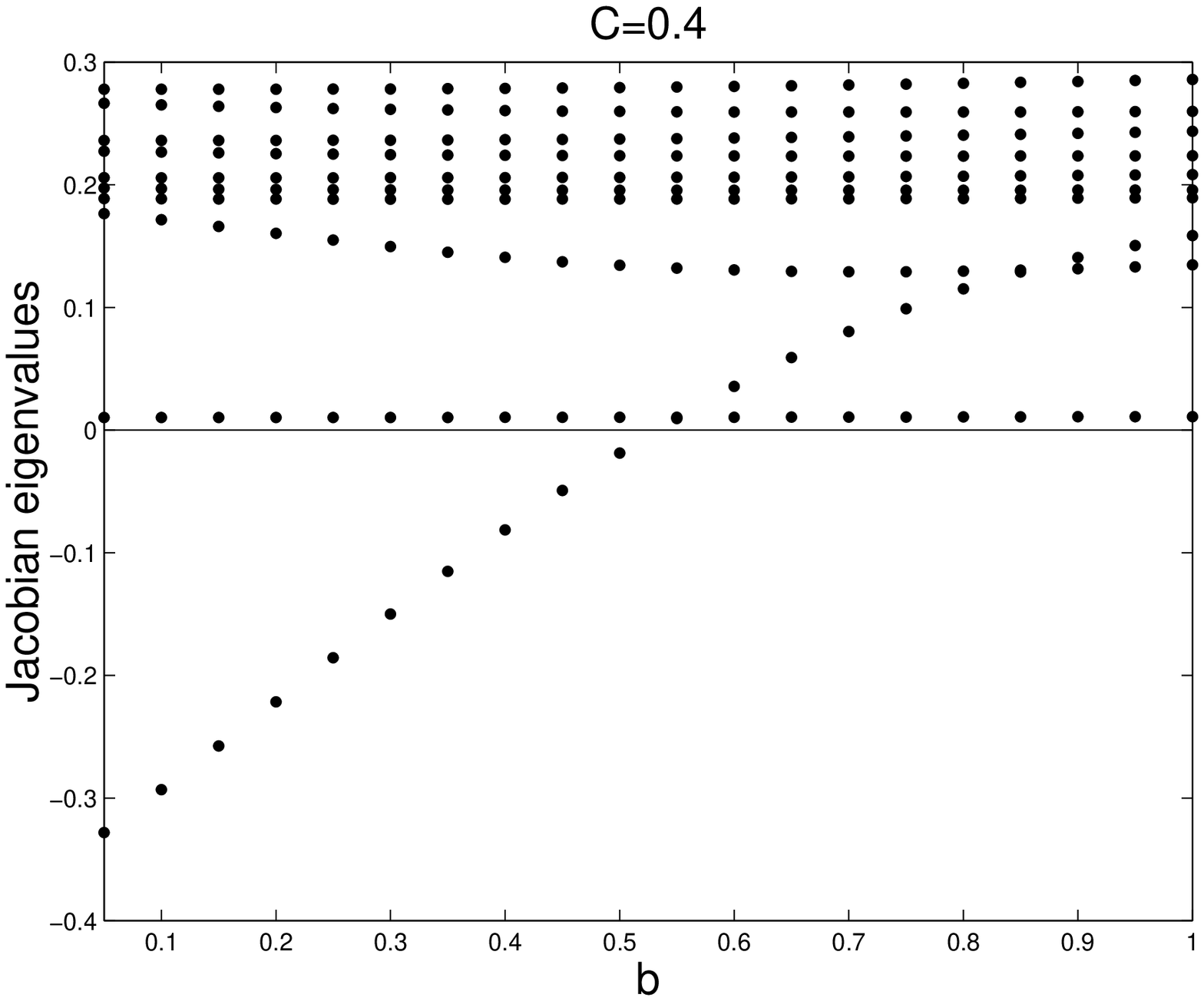}
\end{tabular}
\caption{Dependence of the Jacobian eigenvalues with respect to
$b$. It can be observed that one eigenvalue changes its sign, and
another is constant and close to zero. The first one is
responsible for the bifurcation studied in the text, while the
second one indicates the quasi-stability necessary for
breather mobility \cite{CAGR02}.}%
\label{fig:jacrand}
\end{center}
\end{figure}

The non-existence of forwards vacancy migration can be explained
through a bifurcation. If we analyze the spectrum of the Jacobian
of the dynamical equations (\ref{eq:dyn}) defined by
$\mathcal{J}\equiv\partial_xF(\{x_n\})$, bifurcations can be
detected. A necessary condition for the occurrence of a
bifurcation is that an eigenvalue of $\mathcal{J}$ becomes zero.
Figure \ref{fig:jacrand} shows the dependence of the eigenvalues
closest to zero with respect to $b$ for $C=0.5$ and $C=0.4$. It
can be observed that, in both cases, there is an eigenvalue that
crosses zero in $b\in(0.65,0.70)$ for $C=0.5$ and in
$b\in(0.50,0.55)$ for $C=0.4$. These values agree with the points
where the probability of the jump forward vanishes.

\begin{figure}
\begin{center}
\begin{tabular}{cc}
    \includegraphics[width=\middlefig]{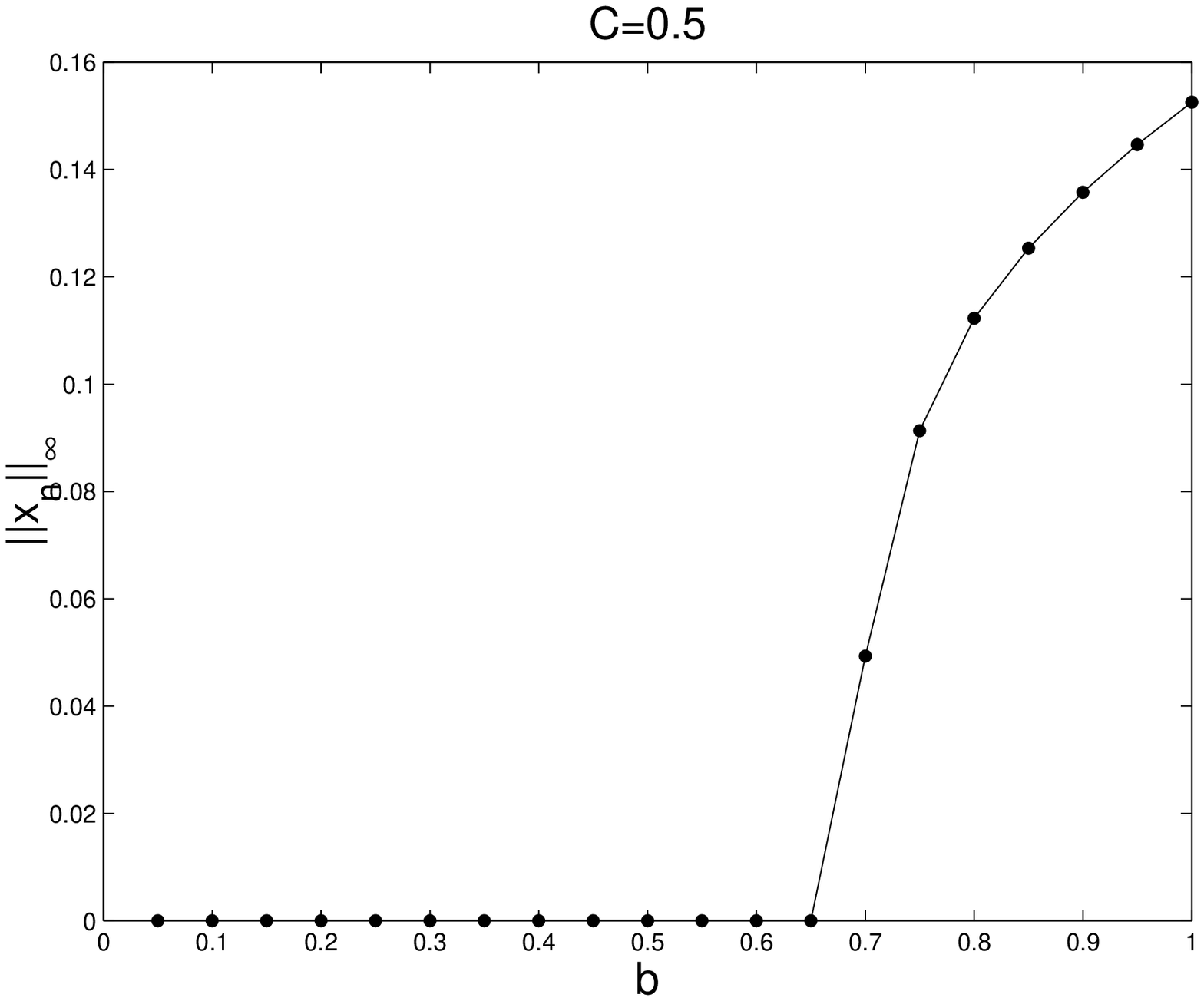} &
    \includegraphics[width=\middlefig]{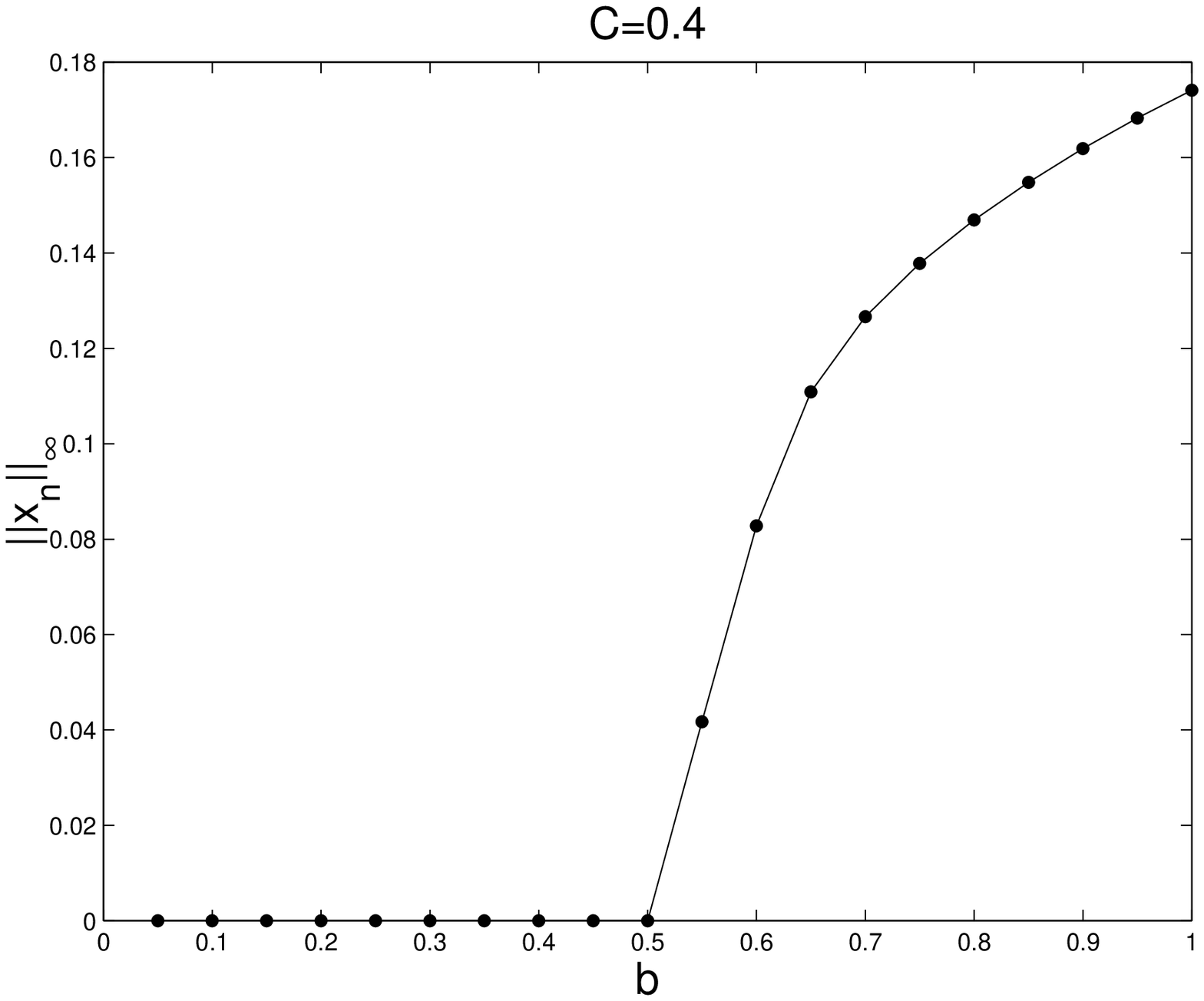}
\end{tabular}
\caption{Amplitude maxima of a vacancy breather versus $b$. It can
be observed that the vacancy breather disappears at the
bifurcation point (see figure \ref{fig:jacrand}).
This is related to the vanishing of the forward movement probability.}%
\label{fig:amprand}
\end{center}
\end{figure}

These bifurcations are related to the disappearance of the
entities we call \emph{vacancy breathers}. They are defined as
breathers centered at the site neighboring to the vacancy, e.g.
the $\nv-1$ or $\nv+1$ sites. It can be observed (figure
\ref{fig:amprand}) that, for $b$ below the bifurcation value,
vacancy breathers do not exist.

\section{Conclusions}

In this paper, we have observed that a moving breather can force a
vacancy defect to move forwards, backwards or let it at its site. We
have also analyzed the influence of the width of the coupling
potential and the coupling strength on the possibility of movement
of a vacancy after the collision with a moving breather. We have
 observed that the width of the potential must
be higher than a threshold value in order that the vacancy can move
forwards.  This behaviour is relevant because experiments
developed in crystals show that the defects are pushed towards the
edges. We have also established that the non--existence of a
breather centered at the sites adjacent to the vacancy is a
necessary condition for the forward vacancy movement.

The incident breathers can be trapped, in the sense that the
energy becomes localized at the vacancy next--neighbors, which
radiate and eventually the energy spreads through the lattice. It
can also be transmitted or reflected. The transmission can only
occur if the vacancy moves backwards. The moving breather always
losses energy but there is not a clear correlation between the
vacancy and breather behaviours.

\ack

The authors acknowledge Prof. F Palmero, from the GFNL of the
University of Sevilla, for valuable suggestions. They also
acknowledge partial support under the European Commission RTN
project LOCNET, HPRN-CT-1999-00163. J Cuevas also acknowledges an
FPDI grant from `La Junta de Andaluc\'{\i}a'.

\newcommand{\noopsort}[1]{} \newcommand{\printfirst}[2]{#1}
  \newcommand{\singleletter}[1]{#1} \newcommand{\switchargs}[2]{#2#1}

\end{document}